\documentclass[11pt]{revtex4-1}
\usepackage{bm}
\usepackage{color,graphicx}
\usepackage{tensor}
\usepackage{amsmath}
\begin{document}
\newcommand{\boldnabla}{\bm{\nabla}}
\newcommand{\boldtheta}{\bm{\theta}}
\newcommand{\angmom}{\bm{\mathcal{J}}}

%%%%%%%%%%%%%%%%%%%%%%%%%%%%%%%%%%%%%%%%%%%%%%%%%%%%%%%%%%%%%%%%%%%%%%%%%%%%%%%%%%%%%%%%%%%%%%%%%%%%%%%%%%%%%%%%%%
%	TITLE, AUTHORS AND DATE
%%%%%%%%%%%%%%%%%%%%%%%%%%%%%%%%%%%%%%%%%%%%%%%%%%%%%%%%%%%%%%%%%%%%%%%%%%%%%%%%%%%%%%%%%%%%%%%%%%%%%%%%%%%%%%%%%%

\title{On single-photon and classical interference}

\author{Stephen M.~Barnett}

\address{School of Physics and Astronomy, University of Glasgow,
Glasgow G12 8QQ, United Kingdom}

%\ead{stephen.barnett@glasgow.ac.uk}

\date{\today}

%%%%%%%%%%%%%%%%%%%%%%%%%%%%%%%%%%%%%%%%%%%%%%%%%%%%%%%%%%%%%%%%%%%%%%%%%%%%%%%%%%%%%%%%%%%%%%%%%%%%%%%%%%%%%%%%%%
%	ABSTRACT AND PACS NUMBERS
%%%%%%%%%%%%%%%%%%%%%%%%%%%%%%%%%%%%%%%%%%%%%%%%%%%%%%%%%%%%%%%%%%%%%%%%%%%%%%%%%%%%%%%%%%%%%%%%%%%%%%%%%%%%%%%%%%

\begin{abstract}
\noindent It has often been remarked that single-photon interference experiments, however complicated, seem to behave very much in the same way as those
performed in the classical regime, using the field generated by a laser.  This observation has the status of being `well-known to those who know it', but perhaps
mysterious to others.  We discuss the reasons underlying the similarity and also some of the limitations of this simple idea.

\end{abstract}

%\pacs{}

\maketitle

% \ioptwocol

%%%%%%%%%%%%%%%%%%%%%%%%%%%%%%%%%%%%%%%%%%%%%%%%%%%%%%%%%%%%%%%%%%%%%%%%%%%%%%%%%%%%%%%%%%%%%%%%%%%%%%%%%%%%%%%%%%
%	INTRODUCTION
%%%%%%%%%%%%%%%%%%%%%%%%%%%%%%%%%%%%%%%%%%%%%%%%%%%%%%%%%%%%%%%%%%%%%%%%%%%%%%%%%%%%%%%%%%%%%%%%%%%%%%%%%%%%%%%%%%

\section{Introduction}

Modern quantum optics and quantum information have encouraged the development of experiments 
using single photons or, to be more precise, in the single-photon regime.  The most popular way of
generating these is to employ spontaneous parametric downconversion, which prepares entangled
photon pairs and uses one of these to herald the presence of the other \cite{Burnham,Friberg,MW}.  This 
heralded photon is then available for the desired experiment or application.  As part of the operation
a laser field is often used to assist with alignment and other practical considerations before employing 
the single photon source, with the measured intensity at various output ports being proportional to
the anticipated detection probabilities in the single-photon regime.  It is noteworthy, moreover, that a
number of important experiments reporting quantum behaviour have used a laser source, sometimes
attenuated, rather than single photons, in the expectation that a single photon would behave in the 
same manner.  A few examples may be in order: quantum key distribution with weak laser pulses 
\cite{PaulSimon,Norbert}, optimal measurements for quantum state discrimination \cite{SarahRev}, 
quantum walks \cite{Christine} and demonstrations of the benefits of indefinite causal order \cite{Jacqui}.

We seek to explain the reason why classical (laser-based) and single-photon interference experiments
are so similar.  Our analysis is based on the behaviour of coherent states of light, which provide the
closest approximation, within quantum theory to a classical radiation field 
\cite{Glauber66,Klauder,GlauberBook}.  The properties of coherent states and, in particular, the way 
in which fields prepared in coherent states interfere provide
simple explanations for the similarity between classical and single-photon interference experiments.

It is important to realise that there are distinct differences between coherent states, even those with 
a very small amplitude, and genuine single photons.  We conclude with a discussion of two of these
differences: photon anti-bunching \cite{QTL3,Alain} and the two-photon Hong-Ou-Mandel effect
\cite{MW,HOM}.

%%%%%%%%%%%%%%%%%%%%%%%%%%%%%%%%%%%%%%%%%%%%%%%%%%%%%%%%%%%%%%%%%%%%%%%%%%%%%%%%%%%%%%%%%%%%%%%%%%%%%%%%%%%%%%%%%%%%%%%%%%%%%%%%%%%%%%%%%%%%%%%%%%%%%%%%%%%%%%%%% 
%%%%%%%%%%%%%%%%%%%%%%%%%%%%%%%%%%%%%%%%%%%%%%%%%%%%%%%%%%%%%%%%%%%%%%%%%%%%%%%%%%%%%%%%%%%%%%%%%%%%%%%%%%%%%%%%%%%%%%%%

\section{Classical and quantum theories of interference}

To address our question we need to make precise a working definition of an interferometer.  We 
define an interferometer as a device comprising only passive linear optical elements, principally 
beam splitters, mirrors and wave-plates, with input ports into which light can be directed and output 
ports which record either photocounts in the single photon regime or photocurrents proportional
to the laser intensity.  It is certainly possible to construct interferometrers that include more
complicated elements, such as nonlinear media, but including these tends to invalidate the link
between laser-based experiments and those utilising single photons.  Examples of suitable
devices include the familiar Mach-Zehnder and Michelson interferometers, but also the famous
Young two-slit experiment.  The components in these devices coherently superpose light from 
interfering modes, redirect light and rotate polarisations.  They do not add photons to the field 
although lossy elements may remove them.

\begin{figure}[h!]
\centering
\includegraphics[width=\textwidth]{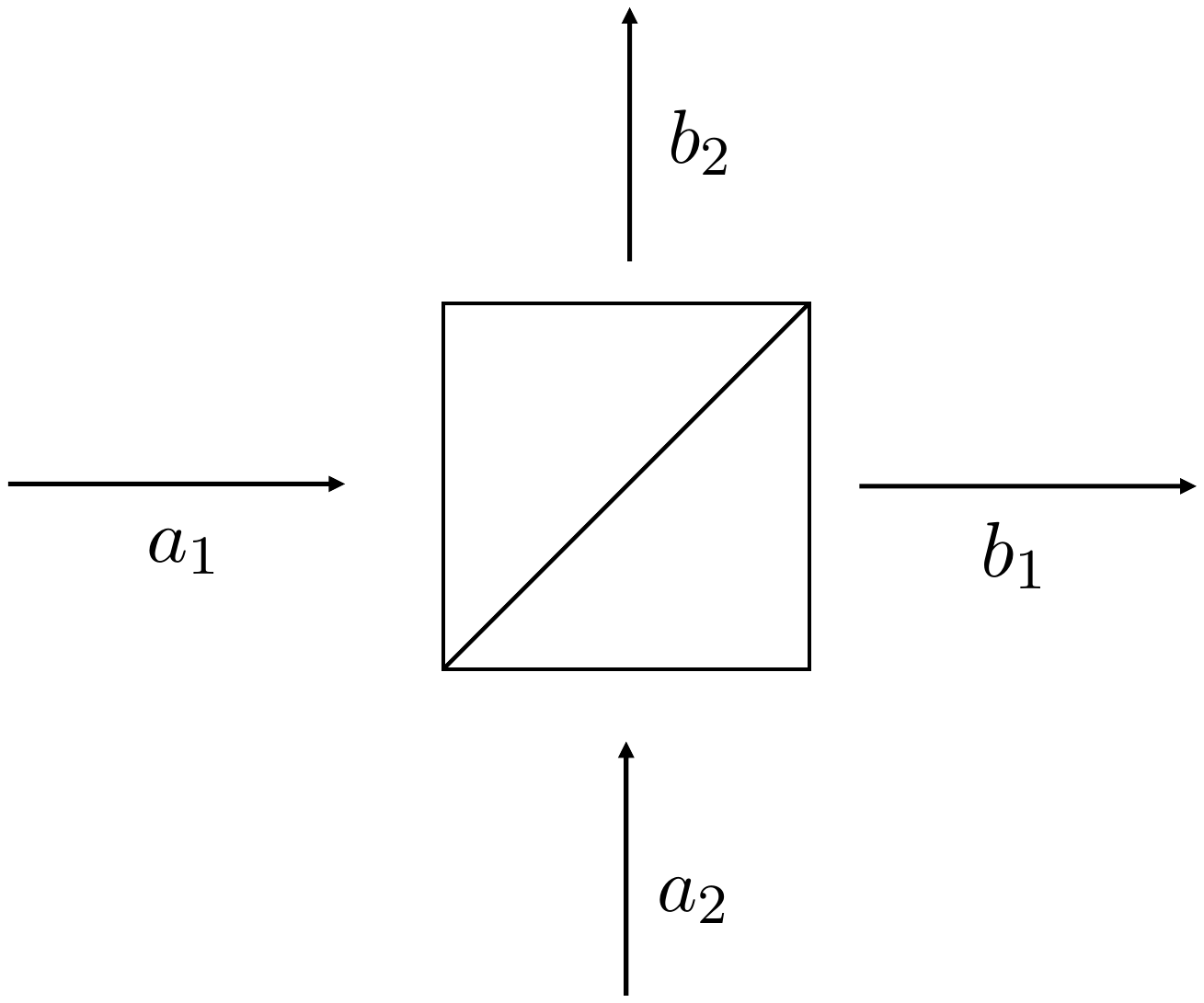}
\caption{Schematic representation of a beam splitter.  In the classical theory, input modes with
complex amplitudes $a_1$ and $a_2$ are superposed to form the output amplitudes $b_1$ and
$b_2$.  In the quantum theory these complex amplitudes are replaced by photon annihilation 
operators.}
\label{Fig1}
\end{figure}

A key element in many practical interferometers is the beam splitter, depicted in Figure~\ref{Fig1}.
In the classical theory this combines two modes with complex amplitudes $a_1$ and $a_2$ to
produce two output modes with amplitudes $b_1$ and $b_2$.  For simplicity we consider a
symmetric beam splitter for which the two sets of amplitudes are related by \cite{QTL3}
\begin{equation}
\label{Eq1}
b_1 = ta_1 + ra_2 \qquad b_2 = ta_2 + ra_1 \, .
\end{equation}
The intensity is proportional to the modulus squared of these amplitudes and it follows that the
total output intensity is proportional to
\begin{equation}
\label{Eq2}
|b_1|^2 + |b_2|^2 = (|t|^2 + |r|^2)(|a_1|^2 + |a_2|^2) + (t^*r + r^*t)(a_1^*a_2 + a_2^*a_1) \, .
\end{equation}
As it stands this expression depends on the relative phase between the amplitudes $a_1$ and $a_2$,
but the values of $t$ and $r$, being fixed properties of the beam splitter,
 cannot depend on this relative phase so we can infer that 
$t^*r + r^*t = 0$, so that $t^*r$ is imaginary.  Conservation of energy then requires that $|t|^2 + |r|^2 = 1$.  In the quantum theory 
we can replace the field amplitudes by photon annihilation operators so that our classical relationship
in Equation~\ref{Eq1} is replaced by the operator form
\begin{equation}
\label{Eq3}
\hat{b}_1 = t\hat{a}_1 + r\hat{a}_2 \qquad \hat{b}_2 = t\hat{a}_2 + r\hat{a}_1 \, .
\end{equation}
The same relationships between $t$ and $r$ as found in the classical theory can now be derived from the 
commutation relations $[\hat{b}_i,\hat{b}_j^\dagger] = \delta_{ij} = [\hat{a}_i,\hat{a}_j^\dagger]$.

\begin{figure}[h!]
\centering
\includegraphics[width=\textwidth]{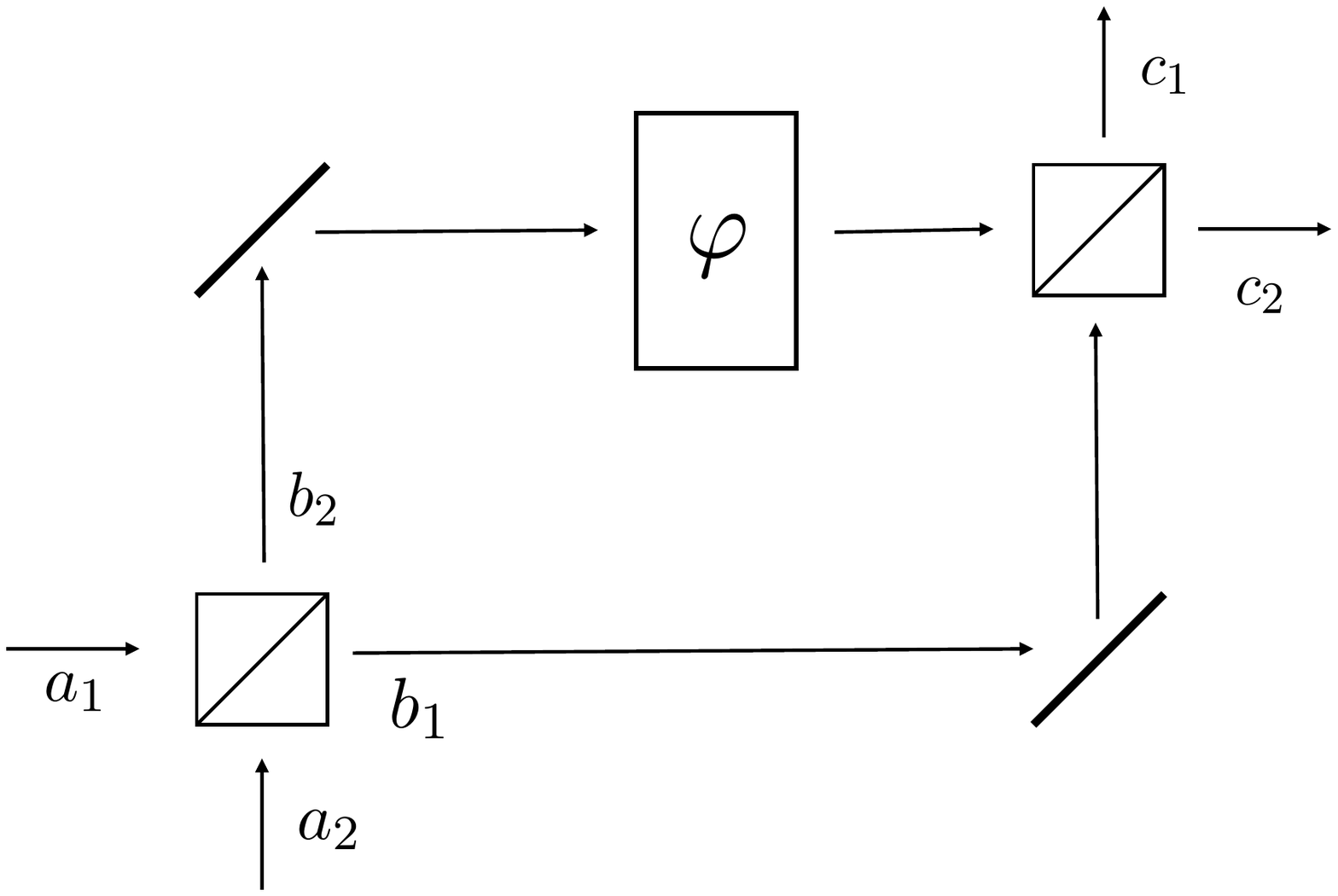}
\caption{Schematic representation of a Mach-Zehnder interferometer.  The input beam splitter is as in Figure~\ref{Fig1}
with transmission and reflection coefficients $t_1$ and $r_1$.  The lengths of the two arms are $L_1$ and $L_2$ so that
propagation along the arms produces the phase shifts $kL_1$ and $kL_2$.  For completeness, we include, also an 
additional element in the upper arm producing an additional phase shift $\varphi$.}
\label{Fig2}
\end{figure}

As an example we consider a simple Mach-Zehnder interferometer as depicted in Figure~\ref{Fig2}.  It is
straightforward to determine the forms of the output amplitudes, $c_1$ and $c_2$ or, in quantum theory,
the output annihilation operators in terms of the input ones.  
\begin{align}
\label{Eq4}
%c_1 &= t_2e^{ikL_1}b_1 + r_2e^{i\varphi}e^{ikL_2}b_2 
%&c_2 &= t_2e^{i\varphi}e^{ikL_2}b_2 + r_2e^{ikL_1}b_1  \nonumber \\
%b_1 &= t_1a_1 + r_1a_2  &b_2 & = t_1a_2 + r_1a_1 \nonumber  \\
%\Rightarrow 
c_1 &= (t_2t_1e^{ikL_1} + r_2r_1e^{i\varphi}e^{ikL_2})a_1 + (t_2r_1e^{ikL_1} + r_2t_1e^{i\varphi}e^{ikL_2})a_2 \nonumber \\
c_2 &= (t_2r_1e^{i\varphi}e^{ikL_2} + r_2t_1e^{ikL_1})a_1 + (t_2t_1e^{i\varphi}e^{ikL_2} + r_1r_2e^{ikL_1})a_2  \,.
\end{align}
If we input light only in mode $a_1$, then the intensities at the outputs will be proportional to the quantities
\begin{align}
\label{Eq5}
|c_1|^2 & = |t_2t_1e^{ikL_1} + r_2r_1e^{i\varphi}e^{ikL_2}|^2|a_1|^2  \nonumber \\
|c_2|^2 & = |t_2r_1e^{i\varphi}e^{ikL_2} + r_2t_1e^{ikL_1}|^2|a_1|^2 \, .
\end{align}
We can think of the quantity $|t_2t_1e^{ikL_1} + r_2r_1e^{i\varphi}e^{ikL_2}|^2$ as the probability that any one 
photon in the input emerges in mode $c_1$ and $|t_2r_1e^{i\varphi}e^{ikL_2} + r_2t_1e^{ikL_1}|^2$ as the
probability that it emerges in mode $c_2$.  We can check this interpretation by turning to the quantum 
description, where we find that the annihilation operators for the output modes are related to those of the 
input by
\begin{align}
\label{Eq6}
\hat{c}_1 &= (t_2t_1e^{ikL_1} + r_2r_1e^{i\varphi}e^{ikL_2})\hat{a}_1 + (t_2r_1e^{ikL_1} + r_2t_1e^{i\varphi}e^{ikL_2})\hat{a}_2 \nonumber \\
\hat{c}_2 &= (t_2r_1e^{i\varphi}e^{ikL_2} + r_2t_1e^{ikL_1})\hat{a}_1 + (t_2t_1e^{i\varphi}e^{ikL_2} + r_1r_2e^{ikL_1})\hat{a}_2  \,.
\end{align}
If we have just a single photon prepared in mode $a_1$ and none in mode $a_2$ then 
$\langle\hat{a}_1^\dagger\hat{a}_1\rangle = 1$ and $\langle\hat{a}_2^\dagger\hat{a}_2\rangle = 0$.  It follows
that the probabilities that the photon leaves the interferometer in mode $c_1$ or $c_2$ are
\begin{align}
\label{Eq7}
\langle\hat{c}_1^\dagger\hat{c}_1\rangle & = |t_2t_1e^{ikL_1} + r_2r_1e^{i\varphi}e^{ikL_2}|^2  \nonumber \\
\langle\hat{c}_2^\dagger\hat{c}_2\rangle & = |t_2r_1e^{i\varphi}e^{ikL_2} + r_2t_1e^{ikL_2}|^2 
\end{align}
respectively.  These are precisely the quantities inferred from the classical description.

In the following section we shall address the question of why the single-photon probabilities arise so
simply in the classical theory and, in doing so, establish that the equivalence is general and may be
applied to any interferometric device.  Before turning to this, it is instructive to recall what three 
others (among many) have written on this point.  We start with the famous quote from Dirac \cite{Dirac}:

{\it Suppose we have a beam of light consisting of a large number of photons split up into
two components of equal intensity.  On the assumption that the intensity of a beam is connected 
with the probable number of photons in it, we should have half the total number of photons 
going into each component.  If the two components are now made to interfere, we should
require a photon in one component to be able to interfere with one in the other.  Sometimes 
these two photons would have to annihilate one another and other times they would have to produce 
four photons.  This would contradict the conservation of energy.  The new theory, which connects
the wave function with probabilities for one photon, gets over the difficulty by making each photon
go partly into each of the two components.  Each photon then interferes only with itself.  Interference 
between two different photons never occurs.}

In the first edition of his book, The Quantum Theory of Light, Loudon writes of Young's interference
experiment \cite{QTL1}:

{\it Photons do not interact with each other, and any interference effects must be sought in the 
process by which each single photon passes from the source to the second screen.  
Quantum-mechanically, the interference occurs between the probability amplitudes for passage 
from source to screen via the two different paths corresponding to the two pinholes.  The 
intensity on the second screen is proportional to the square modulus of the sum of the two
probability amplitudes.  The structure of the quantum-mechanical calculation is the same 
as that of the classical calculation, which is also based of the sum of two amplitudes, and the
two calculations give the same intensity distribution.}

Finally, in Jex's translation of Paul's book, Introduction to Quantum Optics, we find \cite{Paul}:

{\it Let us note that the quantized theory of the electromagnetic field encompasses the particle 
equally well as the wave aspect.  In particular, beamsplitting can be described in such a way
that (in complete correspondence to classical theory) the electric field strength - now described 
by an operator - of the incident wave is decomposed into parts corresponding to the reflected 
wave and the transmitted wave.  We find then the surprising (at least at first glance) result 
that the classical interference pattern is quantum mechanically exactly reproducible independent
of the (perhaps even non-classical) state of the incident light.}

The common point emphasised by each of these accounts is that in interference, as we have
discussed it, single photons within a beam of light behave, individually, in the same manner 
as would a field in the classical theory.

%%%%%%%%%%%%%%%%%%%%%%%%%%%%%%%%%%%%%%%%%%%%%%%%%%%%%%%%%%%%%%%%%%%%%%%%%%%%%%%%%%%%%%%%%%%%%%
%%%%%%%%%%%%%%%%%%%%%%%%%%%%%%%%%%%%%%%%%%%%%%%%%%%%%%%%%%%%%%%%%%%%%%%%%%%%%%%%%%%%%%%%%%%%%%%%%%%

\section{Interfering coherent states}

The key to deriving our desired result, that single-photon interference experiments behave 
precisely as do classical ones, lies in the properties of the coherent states.  For any single mode,
with annihilation and creation operators, $\hat{a}$ and $\hat{a}^\dagger$, the coherent state is
a displaced vacuum obtained by means of a unitary transformation 
\cite{MW,Glauber66,Klauder,GlauberBook,QTL3,Methods}:
\begin{equation}
\label{Eq8}
|\alpha\rangle = \hat{D}(\alpha)|0\rangle = \exp\left(\alpha\hat{a}^\dagger - \alpha^*\hat{a}\right)|0\rangle \, .
\end{equation}
Two properties of the coherent states will be useful to us.  The first of these is that the coherent states are right-eigenstates of the annihilation operator with
eigenvalue $\alpha$:
\begin{equation}
\label{Eq9}
\hat{a}|\alpha\rangle = \alpha|\alpha\rangle \, .
\end{equation}
The second is that the coherent state 
is a superposition of photon number states in the form
\begin{equation}
\label{Eq10}
|\alpha\rangle = e^{-|\alpha|^2/2}\sum_{n=0}^\infty \frac{\alpha^n}{\sqrt{n!}}|n\rangle \, .
\end{equation}

From the eigenvalue property we can determine how coherent state combine when superposed 
on a beam splitter.  In Equation~(\ref{Eq3}) we have, essentially, a Heisenberg picture relation
between the output and input modes for a simple beam splitter.  If we prepare each input mode 
in a coherent state, $|\alpha_1\rangle$ and $|\alpha_2\rangle$ respectively, then it follows that
the state is a right-eigenstate of the output annihilation operators $\hat{b}_1$ and $\hat{b}_2$:
\begin{align}
\label{Eq11}
\hat{b}_1|\alpha_1\rangle|\alpha_2\rangle &= (t\alpha_1 + r\alpha_2)|\alpha_1\rangle|\alpha_2\rangle \nonumber \\
\hat{b}_2|\alpha_1\rangle|\alpha_2\rangle &= (t\alpha_2 + r\alpha_1)|\alpha_1\rangle|\alpha_2\rangle \, .
\end{align}
It follows that, in the Schr\"{o}dinger picture, the states of the output modes are also coherent states,
$|\beta_1\rangle$ and $|\beta_2\rangle$ respectively, where $\beta_1 = t\alpha_1 + r\beta_2$ and
$\beta_2 = t\alpha_2 + r\beta_1$.  Hence the complex amplitudes of our coherent states combine 
in precisely the same manner as do the classical amplitudes.  This conclusion applies to each of our 
passive linear optical elements and, indeed, to any interferometer, however complicated.  This property
was first obtained in the context of coupled oscillators by Glauber \cite{GlauberPL}.  It may be of 
interest to note that this idea can readily be extended to any state of light by writing such states 
as a superposition of coherent states \cite{JanszkySMB}.

Another way to see the relationship between the coherent state amplitudes and the amplitudes in
the classical theory is to adopt the vacuum picture \cite{Mollow,Samson,David,PMRPLK}.  To see how this works
consider a single mode of the field in which the electric field operator has the form
\begin{equation}
\label{Eq12}
\hat{\bf E} = {\bf E}_0 u({\bf r})e^{-i\omega t}\hat{a} + {\bf E}^*_0 u^*({\bf r})e^{i\omega t}\hat{a}^\dagger \, ,
\end{equation}
where $u({\bf r})$ is the spatial mode function.  Let this mode be prepared in the coherent state $|\alpha\rangle$.
We can transform into the vacuum picture by acting on the state with the unitary operator 
$\hat{D}^\dagger(\alpha) = \hat{D}(-\alpha)$, which leaves the field mode in its vacuum state, $|0\rangle$.
To keep the physical situation unchanged we need, also, to transform the operators 
$\hat{O} \rightarrow \hat{D}^\dagger(\alpha)\hat{O}\hat{D}(\alpha)$.  For our electric field operator, this 
transformation produces the operator
\begin{equation}
\hat{D}^\dagger(\alpha)\hat{\bf E}\hat{D}(\alpha) = \hat{\bf E} + 
{\bf E}_0 u({\bf r})e^{-i\omega t}\alpha + {\bf E}^*_0 u^*({\bf r})e^{i\omega t}\alpha^* \, ,
\end{equation}
which is a superposition of the electric field operator and a classical field with an amplitude proportional 
to $\alpha$.  The passive optical elements leave the vacuum field unchanged; zero photons in leads to
zero photons out, and the c-number amplitudes behave as do the amplitudes in the classical
theory.  It follows that a coherent state input into our interferometer behaves in precisely the same
manner as the field amplitude in the classical theory.

The final step is to note that in the number-state expansion of the coherent state, the single-photon
probability amplitude is proportional to $\alpha$, with the amplitude for higher photon numbers 
varying as $\alpha^n$.  It follows that the single-photon probability amplitude evolves on passage
through an interferometer in the same manner as the amplitude $\alpha$ and, therefore, as does the field in the
classical description.  This completes our proof that in an interferometer with single photon input,
the probability amplitude associated with any given path through the device behaves precisely as
does the amplitude of a classical field.

The simplest way to state the equivalence is that the coherent amplitude for a coherent state, $\alpha$,
behaves in the same way as does a classical field amplitude.  For a single photon, $\alpha$ is also the
probability amplitude and the detection probability is proportional to $|\alpha|^2$, as is the intensity in
a classical treatment.

%%%%%%%%%%%%%%%%%%%%%%%%%%%%%%%%%%%%%%%%%%%%%%%%%%%%%%%%%%%%%%%%%%%%%%%%%%%%%%%%%%%%%%%%%%%%%%%%%%%%%%%%%%%%%%%%%%%%%%%%%%%%%%%%
%%%%%%%%%%%%%%%%%%%%%%%%%%%%%%%%%%%%%%%%%%%%%%%%%%%%%%%%%%%%%%%%%%%%%%%%%%%%%%%%%%%%%%%%%%%%%%%%%%%%%%%%%%%%%%%%%%%%%%%%%

\section{Limitations}

It is important not to push the above idea too far.  What we have established is an equivalence between
the probabilities for single detection events at the outputs of a single-photon device with the corresponding
intensity measured in the same interferometer with a laser input.  We consider two quantum interference 
phenomena that are not obtainable classically.  These are photon antibunching and two-photon interference 
in the Hong-Ou-Mandel effect, both of which reveal themselves in the correlations between pairs of 
detectors.

\subsection{Photon antibunching}

If we have just a single photon, then it cannot be detected in two separate detectors.  This means that if we 
place a detector in each of the outputs from our beam splitter in Figure~\ref{Fig1} and send a single photon 
in one of the input arms then we can only get a detection event in one of the two detectors.  The transmission
and reflection coefficients tell us the probability that the photons is transmitted or reflected and then is 
detected in the corresponding detector.  This feature, as we have seen, is reproduced in the classical theory.

In the classical theory if $\alpha$ is the input field amplitude then the probability that a detection event 
occurs in both detectors is proportional to $|t|^2|r|^2|\alpha|^4$.  More precisely, the probability of
detecting light in both output detectors is always at least as big as the product of the probabilities for
detections at the individual detectors, and the anticorrelation between detection events found for a
single photon input cannot be reproduced in the classical theory.

The anticorrelation is crucial to the demonstration of single-photon interference.  Note that the first attempt to 
produce single-photon interference was probably that by Taylor in 1909, who employed a heavily attenuated
light source and very long exposure times (up to three months) in an attempt to determine, through the loss
of visibility, the size of light quanta \cite{Taylor}.  We now know, as shown above, that this experiment could only produce 
interference fringes however faint the input light.  To demonstrate true single-photon interference one needs
a truly single-photon source and a means of verifying this property.  This was achieved nearly eighty years 
after Taylor's work \cite{Alain,GilbertBook}.  A cascade emission provided a herald that a second photon was about
to be emitted and this entered either an interferometer or a beam splitter followed by two detectors.
The observation of high visibility fringes in the interferometer but near perfect anticorrelation in the two 
detectors following the beam splitter finally confirmed the quantum prediction, noted by Dirac, that each
photon interferes only with itself.

\subsection{Two-photon interference}

In a now classic experiment, Hong, Ou and Mandel showed that if a pair of indistinguishable photons 
are incident on a beam splitter, with one photon in each arm, then the output modes tend to produce 
an anticorrelation, with detections in both detectors suppressed.  It is straightforward to confirm this
using the relationships. In Equation~(\ref{Eq3}), between the input and output modes:
\begin{align}
\label{Eq14}
\hat{a}_1^\dagger\hat{a}_2^\dagger|0\rangle &= (t\hat{b}_1^\dagger + r\hat{b}_2^\dagger)(r\hat{b}_1^\dagger + t\hat{b}_2^\dagger)|0\rangle 
\nonumber \\
&= \left[tr(\hat{b}_1^{\dagger 2} + \hat{b}_2^{\dagger 2}) + (t^2 + r^2)\hat{b}_1^\dagger\hat{b}_2^\dagger\right]|0\rangle \, .
\end{align}
Finding a photon in each output mode is suppressed owing to the phase relationship between $t$ and $r$ \cite{footnote}, 
but the probability for finding two photons both in each of the output modes are equal.  If the beam splitter is
balanced so that $|t|^2 = \frac{1}{2} = |r|^2$, then the coefficient of the two photon state 
$\hat{b}_1^\dagger\hat{b}_2^\dagger|0\rangle$ is zero and the two photons are never found in different
detectors \cite{MW,HOM}.

This feature, which like antibunching depends on correlated detections or their absence, cannot be reproduced
in the classical theory or, what comes to the same thing, with coherent states.  To see this it suffices to consider
the two-photon component of input coherent states.  Let the two input modes be prepared in the coherent states
$|\alpha_1\rangle$ and $|\alpha_2\rangle$.  It follows that the two-photon component of the state is
\begin{align}
\label{Eq15}
|\psi\rangle &= \frac{\alpha_1^2}{\sqrt{2}}|2\rangle_1|0\rangle_2 + \alpha_1\alpha_2|1\rangle_1|1\rangle_2 
+ \frac{\alpha_2^2}{\sqrt{2}}|0\rangle_1|2\rangle_2  \nonumber \\
&= \left(\frac{\alpha_1^2}{2}\hat{a}_1^{\dagger 2} + \alpha_1\alpha_2\hat{a}_1^\dagger\hat{a}_2^\dagger 
+ \frac{\alpha_2^2}{2}\hat{a}_2^{\dagger 2}\right)|0\rangle_1|0\rangle_2 \, .
\end{align}
The corresponding (unnormalised) state of the output modes is 
\begin{equation}
\label{Eq16}
|\psi\rangle = \frac{1}{2}\left[(\alpha_1t + \alpha_2r)\hat{b}_1^\dagger + 
(\alpha_1r + \alpha_2t)\hat{b}_2^\dagger\right]^2|0\rangle_1|0\rangle_2 \, .
\end{equation}
The only way to remove the correlations between the two detectors, that is to set the coefficient of 
$\hat{b}_1^\dagger\hat{b}_2^\dagger|0\rangle_1|0\rangle_2 $ equal to zero, is to set either 
$\alpha_1t + \alpha_2r$ or $\alpha_1r + \alpha_2t$ to be zero.  In this case there is perfect
interference between the coherent state amplitudes and all of the light goes into a just one of 
the two output modes.  This is in sharp contrast to the behaviour of two single photons, noted
above, in which correlations between the two detectors are suppressed, but the two photons are 
equally likely to be found in either output mode.

%%%%%%%%%%%%%%%%%%%%%%%%%%%%%%%%%%%%%%%%%%%%%%%%%%%%%%%%%%%%%%%%%%%%%%%%%%%%%%%%%%%%%%%%%%%%%%%%%%%%%%%%%%%%%%%%
%%%%%%%%%%%%%%%%%%%%%%%%%%%%%%%%%%%%%%%%%%%%%%%%%%%%%%%%%%%%%%%%%%%%%%%%%%%%%%%%%%%%%%%%%%%%%%%%%%%%%%%%%%%%%%%%%

\section{Conclusion}

It has long been recognised that single-photon interference experiments behave, essentially, in the same way as those performed 
with far more intense fields such as those generated by a laser.  At the heart of this is the familiar statement by Dirac that ``each 
photon then interferes only with itself" \cite{Dirac}.  Our aim in writing this paper is to explore precisely why this is so.  We have found that,
as is often the case in quantum optics, the coherent states provide a natural way of linking single photon (quantum) behaviour
with that of laser-light (classical) behaviour.  The key idea is that the complex amplitude, $\alpha$, associated with a coherent state 
behaves in precisely the same manner as does the classical, c-number, amplitude in the quantum theory of interference.  Moreover,
this same amplitude is that of the single photon component of the coherent state.

It is important to appreciate that there are limitations to the association between single-photon and classical interference.  In particular
we are limited to devices constructed from passive linear optical elements as these conserve the photon number.  The forms of 
measurement are also restricted; we can compare single-photon detection events only, that is the probabilities for detecting the 
single photon at any given detector, and compare these with the fraction of the intensity recorded in the same output port in the 
classical treatment.  It is by observing the correlations between pairs of detectors (or more) that quantum effects become apparent.
We presented two classic examples of this: anticorrelation at a pair of detectors in the single photon regime and two-photon 
interference in the classical Hong-Ou-Mandel experiment.

We have concentrated exclusively on passive linear optical elements and by doing so have found a very general equivalence 
between the behaviour of single-photon and classical behaviours.  This does not mean, however, that such an equivalence 
will never hold in devices that include nonlinear optical elements.  One important example is a recent realisation of quantum
teleportation in which the state to be teleported is encoded first on a laser field and is then teleported to a distant single photon
\cite{Bereneice}.  One nonlinear optical process produces a pair of entangled photons as is often the starting point in a teleportation 
experiment \cite{QI}.  The comparison step is achieved by a frequency conversion arrangement in which one of the entangled photons
(the local photon) interacts with the laser field to produce a new photon carrying information from both the previously entangled photon and the
laser field.  The teleportation may be viewed as the transfer of the state of one of the laser photons to the distant and 
previously entangled photon.  As with single photon interference, this laser-based teleportation does not share all of the 
features of a single-photon teleportation; it cannot teleport entanglement for example.  It does, nevertheless, transfer
the state of the laser photons to the distant photon and it does this without using knowledge of the state of the laser photons
and this, of course, is the key feature of quantum teleportation.

%%%%%%%%%%%%%%%%%%%%%%%%%%%%%%%%%%%%%%%%%%%%%%%%%%%%%%%%%%%%%%%%%%%%%%%%%%%%%%
%%%%%%%%%%%%%%%%%%%%%%%%%%%%%%%%%%%%%%%%%%%%%%%%%%%%%%%%%%%%%%%%%%%%%%%%%%%%%%%%%%%%%%%%%%%%%%%%%%%%%%%%%%%%%%%%%%

%\ack undefined for revtex4-1. Required for iopart.cls
\begin{acknowledgments}
It is a pleasure to dedicate this work to my old friend and regular collaborator Igor Jex, from whom I have learnt so much and had so much
pleasure in the process.  I thank Jacquiline Romero and 
Jonathan Leach for asking the questions that led me to re-explore the issues discussed in this paper.  I thank the Royal Society for the award of a Research Professorship (RP150122).
\end{acknowledgments}

%%%%%%%%%%%%%%%%%%%%%%%%%%%%%%%%%%%%%%%%%%%%%%%%%%%%%%%%%%%%%%%%%%%%%%%%%%%%%%%%%%%%%%%%%%%%%%%%%%%%%%%%%%%%%%%%%%
%%%%%%%%%%%%%%%%%%%%%%%%%%%%%%%%%%%%%%%%%%%%%%%%%%%%%%%%%%%%%%%%%%%%%%%%%%%%%%%%%%%%%%%%%%%%%%%%%%%%%%%%%%%%%%%%%%

%%%%%%%%%%%%%%%%%%%%%%%%%%%%%%%%%%%%%%%%%%%%%%%%%%%%%%%%%%%%%%%%%%%%%%%%%%%%%%%%%%%%%%%%%%%%%%%%%%%%%%%%%%%%%%%%%%
%	END OF THE DOCUMENT
%%%%%%%%%%%%%%%%%%%%%%%%%%%%%%%%%%%%%%%%%%%%%%%%%%%%%%%%%%%%%%%%%%%%%%%%%%%%%%%%%%%%%%%%%%%%%%%%%%%%%%%%%%%%%%%%%%

\end{document}